\documentclass[12pt]{iopart}
\def\IGNORE#1{}
\usepackage{iopams}
\usepackage{epic,eepic}
\usepackage{psfig}
\def\EE#1{\left\langle{#1}\right\rangle}

\def\d{\mathrm{d}}
\def\mod{\mathrm{mod}\,}
\makeatletter

\DeclareRobustCommand{\binom}{\frbinom{}}
\def\frbinom#1#2#3{{#1{#2\atopwithdelims()#3}}}
\makeatother

\begin{document}

\title[Metastability in Short Range 
Spin Models]{Metastable States in High Order Short-Range 
Spin Glasses}

\author{Viviane M.\ de Oliveira$^{a}$,
        J.\ F.\ Fontanari$^{a}$, and 
        Peter F.\ Stadler$^{b,c}$ \footnote[3]{To
           whom correspondence should be addressed.\\
           Email: {\tt studla@tbi.univie.ac.at},
           Phone: **43 1 4277 52737,
           Fax: **43 1 4277 52793}
}

\address{$^a$Instituto de F{\'\i}sica de S{\~a}o Carlos,
            Universidade de S{\~a}o Paulo,
            Caixa Postal 369, 13560-970 S\~ao Carlos SP, Brazil}

\address{$^b$Institut f{\"u}r Theoretische Chemie, Universit{\"a}t Wien
            W{\"a}hringerstra{\ss}e 17, A-1090 Wien, Austria}

\address{$^c$The Santa Fe Institute, 
            1399 Hyde Park Road, Santa Fe, NM 87501, USA}

\begin{abstract}
The mean number $\EE{\mathcal{N}}$ of metastable states in higher order 
short-range spin
glasses is estimated analytically using a variational method 
introduced by Tanaka and Edwards for very large coordination
numbers. For lattices with small connectivities,
numerical simulations do not show any significant dependence on 
the relative positions of the interacting spins on the lattice,
indicating thus that these systems can be described by a
few macroscopic parameters.
As an extremely anisotropic model we consider the low autocorrelated 
binary spin model and we show through numerical simulations
that its landscape has an exceptionally large number of
local optima.
\end{abstract} 


\maketitle


\section{Introduction}
\label{sect:intro}

The notion of an (adaptive) landscape has proved to be a valuable
concept in theoretical investigations of evolutionary change,
combinatorial optimization, and the physics of disordered systems.
From the mathematical point of view, a landscape consists of three
ingredients: (i) a set $V$ of ``configurations'' which we shall assume
to be finite but very large, (ii) a cost or fitness function
$f:V\to\mathbb{R}$ that evaluates the configurations, and (iii) some
sort of additional geometrical, topological, or algebraic structure
$\mathcal{X}$ on $V$ that allows us to define notions of closeness,
similarity, or dissimilarity among the configurations 
\cite{Stadler:96b,Stadler:99b,Weinberger:90a}. In the simplest
case, $\mathcal{X}$ is an adjacency relation. In this contribution we
shall consider systems consisting of $N$ Ising spins and we shall
assume that two spin configurations $x$ and $x'$ are adjacent when
they differ in the orientation of a single spin, $x_k'=-x_k$. 
We say that $x\in V$ is a {\it local minimum} of the landscape $f$ if
$f(x)\le f(y)$ for all adjacent configurations (neighbors) $y$ of
$x$. The use of $\le$ instead of $<$ is conventional
\cite{Kern:93,Ryan:95}; it does not make a significant difference for
spin glass models. Local maxima are defined analogously. The number
$\mathcal{N}$ of local optima in a landscape may serve as a measure
for the landscape's ruggedness \cite{Palmer:91}.

Alternatively, ruggedness can be measured by means of correlation
functions such as $r(s)$, defined as the autocorrelation function 
of the ``time series'' $f(x_t)$ sampled along an unbiased random
walk of $s$ steps on the configuration space \cite{Weinberger:90a}. 
Explicitly, $r(s)$ can be represented as a quadratic form 
\begin{equation}
  r(s) = (\tilde f,(\mathbf{D}^{-1}\mathbf{A})^s \tilde f) \big/ 
   (\tilde f,\tilde f)
\end{equation}
where $\mathbf{A}$ is the adjacency matrix of the configuration space
$(V,\mathcal{X})$, $\mathbf{D}$ is the diagonal matrix of its vertex
degrees, i.e., $\mathbf{D}_{xx}$ is the number of neighbors of each
configuration, $\tilde f(x)=f(x)-\overline{f}$, and
$\overline{f}=|V|^{-1}\sum_x f(x)$ \cite{Stadler:96b}. It is not hard to
verify that $r(s)$ is an exponential function if and only if $\tilde f$ is
an eigenvector of $\mathbf{A}$. Such landscapes have been termed {\em
elementary}. The importance of elementary landscapes derives in part from
the fact that all Ising spin models
\begin{equation}\label{H}
\mathcal{H}\left(x\right) = 
\sum_{\left(i_1<i_2<\ldots<i_p\right)}
J_{i_1 i_2 \ldots i_p} \, x_{i_1} x_{i_2} \ldots x_{i_p}
\end{equation}
with a fixed interaction order $p$ are elementary, with an eigenvalue
$N-2p$ that depends only on $p$ and not on the details of the index set
$\left(i_1<i_2<\ldots<i_p\right)$ of non-vanishing spin interactions.
In the case that the couplings $J_{i_1\ldots i_p}$ are statistically
independent, Gaussian distributed random variables, Eq.(\ref{H})
defines Derrida's $p$-spin Hamiltonian \cite{Derrida:80a}, which
for $p=2$ reduces to the well-known SK model \cite{Sherrington:75}.

The information contained in $r(s)$ is conveniently further condensed
into the {\em correlation length} $\ell = \sum_{k=0}^\infty r(s)$. For
elementary Ising spin models we obtain immediately $\ell=N/(2p)$, see
e.g.\ \cite{Weinberger:93a,Stadler:94a}. It would appear that
$\mathcal{N}$ and $\ell$ are two sides of the same coin, and hence we
would expect a close connection between the two measures. Indeed,
for Derrida's $p$-spin Hamiltonian 
the expected number of local optima $\EE{\mathcal{N}}$ 
scales like $ \exp \left ( \alpha N \right ) $ with $\alpha$ increasing
from $0.199$ for $p=2$ to $\ln 2 \approx 0.692$ for $p \rightarrow \infty$
\cite{Gross:84}. 
This increase of $\alpha$, and hence of the number of local optima, 
matches the decrease of the correlation length $\ell$ with increasing $p$ .

Although for random landscapes
(Hamiltonians with disorder) it is often desirable to determine
$\EE{\ln\mathcal{N}}$,  in most cases one has
$\EE{\ln\mathcal{N}}=\ln\EE{\mathcal{N}}$ (a notable exception is the
linear spin chain \cite{Derrida:86}). The reason for this 
equality is that the overlap between two randomly chosen
metastable states vanishes with probability one and so the
replica approach needed to evaluate the average of 
$ \ln \mathcal{N} $  reduces to the annealed approximation,
which takes the average directly on $\mathcal{N}$ \cite{Roberts}.
Of course, this is no longer true if one considers 
specific classes of metastable states (e.g.\ those possessing
a given energy density)  or if one adds an external magnetic
field to the Hamiltonian (\ref{H}) \cite{Oliveira:97a}.

In \cite{Stadler:96b,Stadler:99b} the notion of an {\em isotropic} random
landscape was introduced as a ``statistically symmetric model'', that is,
as a random landscape with a covariance matrix
$\mathbf{C}_{xy}=\EE{f(x)f(y)}-\EE{f(x)}\EE{f(y)}$ that shares the
symmetries of the underlying configuration space.  An Ising spin glass is
isotropic if and only if all interaction coefficients $J_{i_1\ldots i_p}$ 
are uncorrelated with mean $\EE{J_{i_1\ldots i_p}}=0$, and 
those $J_{i_1\ldots i_p}$ that belong
to a common interaction order have the same variance,
$\EE{J^2_{i_1\ldots i_p}}=\sigma^2(p)$ \cite{Stadler:99b}. 
In other words, the
elementary isotropic Ising models are exactly the $p$-spin Hamiltonians
with infinite-range interactions. It
is argued at length in \cite{GarciaPelayo:97a,Stadler:99g} that isotropy
can be interpreted as a maximum entropy condition. The properties of the
infinite-range $p$-spin Hamiltonians are obviously determined exclusively 
by the number of
spins $N$ and the interaction order $p$ or, equivalently, by the
correlation length $\ell$.

Short-range spin models, in which only a small fraction of all
$\binom{N}{p}$ possible $p$-ary spin interactions contribute to the
Hamiltonian, i.e., $\big\langle J^2_{i_1i_2\dots i_p}\big\rangle=0$ for
most $p$-ary spin patterns $(i_1<i_2<\dots<i_p)$, deviate significantly
from isotropy. For the SK model ($p=2$) 
a slightly larger number of
local optima has been found \cite{Tanaka:80a,Bray:81} than for the
infinite-range case \cite{Bray:80}. The deviation is proportional to
$1/z$, where $z$ is the number of the nearest-neighboring points
in a hypercubic lattice of dimension $d= z/2$.

In this contribution we show that an analogous effect is at work in 
higher order spin glasses. In addition, numerical simulations for small 
values of connectivities do not show a 
substantial dependence on the patterns of interacting spins. 
The rest of this paper is organized in the 
following way.
In section \ref{sect:SR4spin} we generalize the variational method
of Tanaka and Edwards \cite{Tanaka:80a} to estimate
the $1/z$ corrections to the expected number of metastable states 
of Derrida's $p$-spin model. The results of numerical simulations
of fourth-order spin glasses in  two- and three-dimensional
cubic lattices with small connectivities are discussed in
section \ref{sect:num}. In section \ref{sect:LABSP} we investigate
the metastable states of  a highly frustrated one-dimensional
spin model without explicit disorder, namely, 
the low autocorrelated binary string problem \cite{Bernasconi:87}.
Finally, in section \ref{sect:disc} we present some concluding
remarks.

\section{Short-Range $p$-spin Models}
\label{sect:SR4spin}

The coupling strengths $J_{i_1\ldots i_p}$ in Eq.(\ref{H}) are
modeled as statistically independent random variables with the Gaussian
distribution
\begin{equation}\label{prob}
\mathcal{P}\left(J_{i_1i_2\ldots i_p}\right)=
\sqrt{\frac{z^{p-1}}{\pi p!}}\exp\left[
-\frac{\left(J_{i_1i_2\ldots i_p}\right)^2 z^{p-1}}{p!}\right]
\Theta[(i_1<i_2< \ldots< i_p)] 
\end{equation}
where $\Theta[(i_1<i_2<\ldots< i_p)]=1$ if $(i_1<\ldots<i_p)$ is a valid
interaction pattern and $0$ otherwise.  Since in practice it is not
feasible to consider a fixed interaction pattern, we consider 
all interaction patterns with given fixed coordination number $z$. In other
words, we sum over all ways of choosing the $p-1$ spins among the
$z$ allowed ones.

As an immediate consequence of Eq.(\ref{H}) the energy cost of
flipping the spin $x_j$ is $\delta \mathcal{H} = 2 \lambda_j$, where
\begin{equation}\label{delta}
\lambda_j = \sum_{\left(i_2<\ldots<i_p\right)} 
J_{j\,i_2\ldots i_p} x_j\, x_{i_2} \ldots x_{i_p}
\end{equation}
is the stability of $x_i$. Hence any state $x$ that satisfies 
\begin{equation}
\lambda_i \geq 0 \hskip 2cm  \forall i 
\end{equation}
is a local minimum of the random landscape defined in Eq.(\ref{H}).
Thus the number of local minima can be written as 
\begin{equation}\label{n1}
\mathcal{N}  =  \mbox{Tr}_x \prod_j \int_0^\infty \d\lambda_j
\, \delta \left( \lambda_j -  \sum_{\left(i_2<\ldots<i_p\right)} 
J_{j\, i_2\ldots i_p} x_j\, x_{i_2} \ldots x_{i_p} \right) 
\end{equation}
where $\mbox{Tr}_x$ denotes the summation over the $2^N$ spin
configurations 
and $\delta\left(x\right)$ is the Dirac delta function.

In the following we will calculate analytically the expected number of
metastable states $\langle\mathcal{N}\rangle$ in the limit of large $N$ and
$z$ with $ N \gg z$. Here $\langle\ldots\rangle$
stands for an average over the coupling strengths in all
possible interaction patterns with fixed coordination number $z$.
Using the integral representation of the delta function, the
average over the couplings as well as the summation over the spin
configurations can be easily performed \cite{Oliveira:97a}, yielding
\begin{eqnarray}\label{np0}
\langle\mathcal{N}\rangle &=& \prod_i \int_0^\infty d \lambda_i 
\int_{-\infty}^\infty \frac{\d \phi_i}{\pi} \exp\left({\bf i}
\lambda_i \phi_i\right) \nonumber \\
& & \quad \times  \exp\left[-\frac{p!}{4 z^{p-1}}
\sum_{\left(i_1<i_2<\ldots<i_p\right)} 
\left(\phi_{i_1} + \phi_{i_2} + \ldots + \phi_{i_p} \right)^2
\right]\,.
\end{eqnarray}

Clearly, the expansion of the quadratic term in the 
argument of the exponential function will lead to interactions 
terms of second order in the auxiliary fields $\phi_i$.
More precisely, using 
\begin{eqnarray}\label{res1}
 \frac{p!}{z^{p-1}} \sum_{\left(i_1<\ldots<i_p\right)}
\phi_{i_1}^2 &  = & \frac{1}{z^{p-1}}\, z \left(z-1\right)\ldots
\left(z-p+2\right) \sum_i \phi_i^2 \nonumber \\
& = &
\left[\, 1-\frac{1}{2z}\left(p-1\right)\left(p-2\right) +
         \mathcal{O}\left(z^{-2}\right)
\right]  \sum_i \phi_i^2 
\end{eqnarray}
and
\begin{eqnarray}\label{res2}
\frac{p!}{z^{p-1}} \sum_{\left(i_1<\ldots<i_p\right)}
\phi_{i_1} \phi_{i_2}  & = & \frac{2}{z^{p-1}}\,
\left(z-1\right)\left(z-2\right)\ldots\left(z-p+2\right)
\sum_{\left ( i<j \right )} \phi_i \phi_j \nonumber \\
& = & \left[\frac{2}{z} + \mathcal{O}\left(z^{-2}\right)\right]
      \sum_{\left(i<j\right)} \phi_i \phi_j\,,
\end{eqnarray}
we write Eq.(\ref{np0}) as 
\begin{equation}\label{np}
\langle\mathcal{N}\rangle = \int_{-\infty}^\infty \prod_i \d\phi_i
\, D\left(\phi_i\right) \exp\left[ -\frac{p\left(p-1\right)}{2z} 
\sum_{\left(i<j\right)} \phi_i \phi_j \right]
\end{equation}
where
\begin{equation}
D\left(\phi_i\right) = 
\left[1+\frac{p\left(p-1\right)\left(p-2\right)}{8z}
 \phi_i^2 + \mathcal{O}\left( z^{-2}\right )\right] 
 \,
 \int_0^\infty \frac{\d\lambda}{\pi} \mathrm{e}^{-\mathbf{i}
 \lambda\phi_i-p\phi_i^2/4} 
\end{equation}
is the field's weight function. As mentioned before, Eqns.(\ref{res1}) and
(\ref{res2}) follow from the sum over all possible interaction patterns
with coordination number $z$. More precisely, for each site $i$ there are
precisely $z$ sites $k$ such that $(i,k,i_3,\dots,i_p)$ is a valid
interaction pattern for some choice of $i_3,\dots,i_p$. This notion of site
connectivity is independent of the lattice dimensionality; In fact, it is
not necessary to assume that the sites are arranged on a lattice at all:
Eq.(\ref{np}) remains valid as long as $z$ is site independent.  Hence we
will refer to $z$ simply as the {\em connectivity} of our model.
  
To proceed further we must evaluate the integrals in Eq.(\ref{np}) taking
the care to collect all terms of first order in $1/z$. In particular, note
that $\sum_{(i<j)}1\, =Nz/2$. This can be achieved through an ingenious
variational method introduced by Tanaka and Edwards in their analysis of
the case $p=2$ \cite{Tanaka:80a}. The idea is to add an auxiliary
single-particle term to the effective Hamiltonian of Eq.(\ref{np}) which is
then rewritten as
\begin{equation}\label{nt1}
\langle\mathcal{N}\rangle =
\left[\Phi\left(t\right)\right]^N \, \left\langle \exp\left[ 
 -\frac{p\left(p-1\right)}{2z} 
  \sum_{\left(i<j\right)} \phi_i \phi_j - \mathbf{i}\;t\;p^{1/2} 
  \sum_i \phi_i
\right]\, \right\rangle_{\!t},
\end{equation}
where the average $\langle \ldots \rangle_t$ is defined by
\begin{equation}
\langle\ldots\rangle_t = \frac{ \int \prod_i \d\phi_i
\, D\left(\phi_i\right) \left(\ldots\right)
\exp\left(\mathbf{i}\;t\;p^{1/2}\;\sum_i\phi_i\right)}
{\int\prod_i \d\phi_i \, D\left(\phi_i\right) 
\exp\left(\mathbf{i}\;t\;p^{1/2} \sum_i \phi_i \right)}
\end{equation}
and
\begin{eqnarray}\label{Phi}
\Phi\left(t\right) & = & \int_{-\infty}^\infty
\d\phi\, D\left(\phi\right) \exp \left( \mathbf{i}\;t\;p^{1/2}\phi \right)
\nonumber \\
& = & \mbox{erfc}\left(-t\right)
+ \frac{1}{z}\frac{t}{\sqrt{4\pi}} 
\left(p-1\right)\left(p-2\right)\mathrm{e}^{-t^2} 
+ \mathcal{O}\left(z^{-2}\right)\,.
\end{eqnarray}
For example, we find
\begin{eqnarray}
\mathbf{i}\;p^{1/2}\;\langle\phi_k\rangle_t & =  &\Phi'(t)/\Phi(t)
\label{mom1}  \\
 -p\;\langle\phi_k^2\rangle_t & = & \Phi''(t)/\Phi(t) \label{mom2} 
\end{eqnarray}
for $k=1, \ldots,N$.  Here $t$ is a variational parameter which will be
determined so as to maximize $\langle \mathcal{N} \rangle$. At this point
we can note that the $1/z$ expansion is valid provided that $z \gg p^2$.

As usual, the average in Eq.(\ref{nt1}) can be evaluated through the
cumulant expansion. In particular, we assume that only the first cumulant
contains terms of zeroth order in $1/z$ and it is solely these terms that
determine the value $t=t_m$ that maximizes $\langle\mathcal{N}\rangle$
\cite{Tanaka:80a}.  Of course, this assumption must be verified {\it a
posteriori} through the explicit calculation of the higher order cumulants.
Evaluating the first cumulant we find that $t_m$ maximizes the expression
\begin{equation}
\ln\Phi(t) - \frac{p\left(p-1\right)}{4}
  \langle\phi_k\rangle_t^2 - 
  \mathbf{i}\;t\;p^{1/2}\langle\phi_k\rangle_t 
\end{equation}
and so it is given by the solution of the equation
\begin{equation}\label{tm}
t_m = \frac{1}{2} \left(p-1\right)
      \mathbf{i}\;p^{1/2}\;\langle\phi_k\rangle_{t_m} .
\end{equation}

The usefulness of the variational approach becomes transparent 
only when we rewrite Eq.(\ref{nt1}) replacing $t$ by $t_m$
and rearranging the terms: 
\begin{equation}\label{nt2}
\langle\mathcal{N}\rangle  =
  \exp\left[N\ln\Phi\left(t_m\right)
     +N\frac{p\left(p-1\right)}{4}\langle\phi_k\rangle_{t_m}^2
     \right]\,\times\,
\left\langle\exp\left(\Xi\right)\right\rangle_{t_m}
\end{equation}
where
\begin{equation}
\Xi = - \frac{p\left(p-1\right)}{2z} 
\sum_{\left(i<j\right)} 
\left(\,\phi_i - \langle\phi_i\rangle_{t_m}\,\right)
\left(\,\phi_j - \langle\phi_j\rangle_{t_m}\,\right)\,.
\end{equation}
Since $\left\langle\Xi\right\rangle_{t_m} = 0$,
it is  now straightforward to evaluate 
the higher order cumulants in the cumulant expansion
\begin{equation}\label{ex_c}
\left\langle\exp\left(\Xi\right)\right\rangle_{t_m}
= \exp\left(\sum_{n=1}^\infty \frac{1}{n!} 
\left\langle\Xi^n\right\rangle_{t_m;\mathrm{c}}\right)\,.
\end{equation}
In particular, we find
\begin{eqnarray}
\left\langle\Xi^1\right\rangle_{t_m;\mathrm{c}} & = &  0 , \\
\left\langle\Xi^2\right\rangle_{t_m;\mathrm{c}} & = 
& N \frac{p^2\left(p-1\right)^2}{8z} ~
\left\langle\left(\,\phi_k-\langle\phi_k\rangle_{t_m}\,\right)^2
\right\rangle_{t_m}^2 , \\
\left\langle\Xi^3\right\rangle_{t_m;\mathrm{c}} & =
&  -N \frac{p^3\left(p-1\right)^3}{16z^2} ~
\left\langle\left(\,\phi_k-\langle\phi_k\rangle_{t_m}\,\right)^3
\right\rangle_{t_m}^2 \,.
\end{eqnarray}
More generally, we can easily show that
$\left\langle\Xi^n\right\rangle_{t_m;\mathrm{c}}$ is of order
$z^{1-n}$. Therefore, to keep terms up to $1/z$ we need to consider only the
second cumulant in Eq.(\ref{ex_c}).  Moreover, the leading terms of the
$1/z$ expansion (i.e., the terms of zeroth order in $1/z$) are those
outside the average symbol in Eq.(\ref{nt2}).  Clearly, these results
remain unchanged by the fact that for $p>2$ the averages
$\langle\phi_k^n\rangle_{t_m}$ also contribute to the $1/z$ corrections,
see Eqns.(\ref{Phi}-\ref{mom2}). Hence,
\begin{eqnarray}\label{nt3}
\frac{1}{N}\ln\langle\mathcal{N}\rangle  & = &
\ln\Phi\left(t_m\right) + \frac{p\left(p-1\right)}{4}
                          \langle\phi_k\rangle_{t_m}^2 \nonumber \\
& & + \frac{p^2\left(p-1\right)^2}{16z} ~
      \left(\left\langle\phi_k^2\right\rangle_{t_m}-
            \langle\phi_k\rangle_{t_m}^2 \right)^2 +
      \mathcal{O}\left(z^{-2}\right)\,.
\end{eqnarray}

The next step is to separate the contributions of zeroth and first order in
$1/z$. To achieve this we note that $t_m$ is determined only by the zeroth
order term in the $1/z$ expansion so that Eq.(\ref{tm}) reduces to
\begin{equation}\label{t0}
t_m = \frac{p-1}{\sqrt{\pi}}\,
\frac{\exp\left(-t_m^2\right)}{\mathrm{erfc}\left(-t_m\right)}\,.
\end{equation}
Inserting this result in Eqns.(\ref{Phi}), (\ref{mom1}) and (\ref{mom2}),
we rewrite Eq.(\ref{nt3}) as
\begin{equation}\label{final}
\lim_{N\to\infty} \frac{1}{N}\ln\langle\mathcal{N}\rangle = \alpha =
   \alpha_0 + \frac{1}{z}\alpha_1 + \mathcal{O}\left(z^{-2}\right)
\end{equation}
with
\begin{eqnarray}
\alpha_0 &=& \ln\left[\mathrm{erfc}\left(-t_m\right)\right]-\frac{t_m^2}{p-1}
  \label{a0} \\
\alpha_1 &=& p^2 t_m^2\,\left[\frac{t_m^2}{\left( p-1 \right )^2}
+ \frac{p-2}{2p^2} \right]\, . \label{a1} 
\end{eqnarray}

\begin{table}
\begin{center}
\caption{Values of $\alpha_0$ and $\alpha_1$ obtained from Eqns.(\ref{t0}),
(\ref{a0}), and (\ref{a1}).}
\label{tab:alpha}
\medskip
\begin{tabular}{|l|c|c|c|c|c|c|}
\hline
$p$         & 2        & 3 	& 4 	 & 5      & 6      & 10 \\
\hline
$\alpha_0$  & 0.1992   & 0.2956 & 0.3552 & 0.3965 & 0.4273 & 0.5001\\ 
$\alpha_1$  & 0.0656   & 0.3393 & 0.7789 & 1.3411 & 1.9985 & 5.2894\\
\hline
\end{tabular}
\end{center}
\end{table}

In Table~\ref{tab:alpha} we summarize the values of $\alpha_0$ and
$\alpha_1$ obtained from a numerical solution of Eq.(\ref{t0}).  For $p
\rightarrow \infty$ we find $t_m \approx \sqrt{\ln p}$ and so $\alpha_0
\rightarrow \ln 2$ while $\alpha_1$ diverges like $p \ln p$. Of course,
$\alpha_1/z \ll 1$ since, as mentioned before, the $1/z$ expansion is
consistent only if $z \gg p^2$.  We note that our results for $p=2$ are in
agreement with those of ref.\ \cite{Tanaka:80a}; the values of $\alpha_0$
agree with the TAP solution
\cite{Gross:84}, and with a numerical survey \cite{Stadler:96a}.

\section{Simulations}
\label{sect:num}

The analytical results in section \ref{sect:SR4spin} are valid for very
large connectivities, $z\gg p^2$, and for superimposed interaction
patterns. Actually, the later restriction amounts to consider all
interactions patterns as equivalent, in the sense that they have on the
average the same number of local minima.  In order to investigate the
effect of distinct interaction patterns we consider in this section a
variety of 4-spin models with small connectivities.

\begin{figure}[hbt]
\par
\centerline{\psfig{file=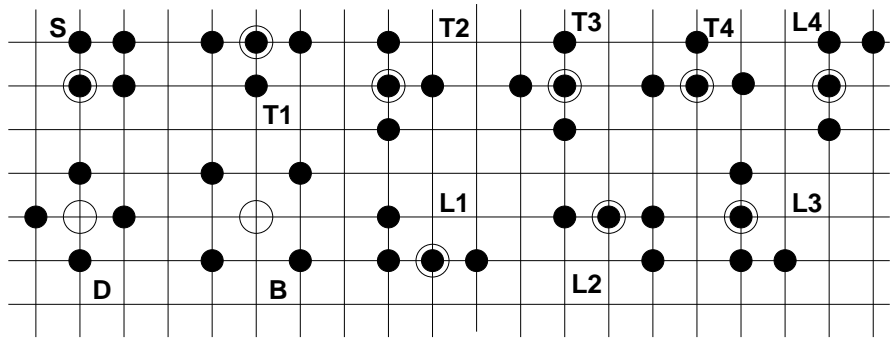,width=0.75\linewidth}}
\par
\caption{Interaction patterns of various short-range $4$-spin models. For
each grid position (indicated by the open circle) a non-zero contribution to
the energy depends on the four spins indicated by the black circles.}
\label{fig:pattern}
\end{figure}

For the sake of computational simplicity we consider only translation
invariant interaction patterns $\mathsf{P}$ on two- and three-dimensional
cubic lattices with periodic boundary conditions. We identify each spin by
its $d$-dimensional lattice coordinates and take all indices modulo the
lattice sizes $m_1$ through $m_d$. In two dimensions, for instance,
translation invariance means
$(i_1,j_1;i_2,j_2;i_3,j_3;i_4,j_4)\in\mathsf{P}$
if and only if\\
$(i_1+p,j_1+q;i_2+p,j_2+q;i_3+p,j_3+q;i_4+p,j_4+q)\in\mathsf{P}$
for all integers $p \, \mod m_1$ and $q \, \mod m_2$. 
The Hamiltonian (\ref{H}) becomes
\begin{equation}\label{H4}
  \mathcal{H}(x) = \sum_{(i_1j_1,i_2j_2,i_3j_3,i_4j_4)\in\mathsf{P}}
     J_{i_1j_1;i_2j_2;i_3j_3;i_4j_4}
     x_{i_1j_1}x_{i_2j_2}x_{i_3j_3}x_{i_4j_4}
\end{equation}
in this example.  A variety of interaction patters on a square lattice,
compiled in Figure~\ref{fig:pattern} and named according to the number and
type of combinations (patterns) of spins that contribute to the Hamiltonian
(\ref{H4}), have been used.  In addition, we have considered three patterns
on a cubic lattice: (i) The ``positive orthant'', coupling each spin
$x_{i,j,k}$ with its lattice neighbors along $x_{i+1,j,k}$, $x_{i,j+1,k}$
and $x_{i,j,k+1}$, (ii) the pattern {\sf S} restricted to a fixed plane,
i.e., $x_{i,j,k}$ coupled with $x_{i,j+1,k}$, $x_{i,j,k+1}$, and
$x_{i,j+1,k+1}$, and (iii) the couplings with the neighbors in all eight
orthants.

\begin{table}
\begin{center}
\caption{Estimated values of $\alpha$ for various short-range 
$4$-spin models.\\ For each pattern we list the number $X$ of non-zero
interaction coefficients per spin, the site connectivity $z$, the
best numerical estimate for $\alpha$ and standard deviation estimated
from the linear regression.}
\label{tab:psp}
\medskip
\begin{tabular}{|ll|cc|rr|}
\hline
Model     & Pattern           & $X$ & $z$ & $\alpha$ & sdev  \\
\hline
{\tt q}   & {\sf S}           &  1  & 8  & 0.4056 & 0.0027 \\ 
{\tt qdq} & {\sf D}           &  1  & 8  & 0.4058 & 0.0017 \\
{\tt q1Tq}& {\sf T1}          &  1  & 10 & 0.3820 & 0.0024 \\
{\tt q1q} & {\sf B}           &  1  & 8  & 0.3885 & 0.0266 \\
\hline
{\tt q2q} & {\sf S,D}         &  2  & 12 & 0.3830 & 0.0049 \\
{\tt qpq} & {\sf T3,T4}       &  2  & 12 & 0.3830 & 0.0013 \\
\hline
{\tt q3q} & {\sf S,D,B}       &  3  & 16 & 0.3695 & 0.0014 \\
{\tt q4q}  & {\sf T1-T4}       &  4  & 12 & 0.3807 & 0.0009 \\
{\tt q5q} & {\sf T1-T4,D}     &  5  & 12 & 0.3744 & 0.0012 \\
{\tt q6q} & {\sf T1-T4,D,B}   &  6  & 16 & 0.3746 & 0.0012 \\
{\tt q8q} & {\sf T1-T4,L1-L4} &  8  & 20 & 0.3744 & 0.0014 \\
{\tt qAq} & {\sf T,L,D,B}     & 10  & 24 & 0.3757 & 0.0026 \\
\hline
{\tt c}   & octant            &  1  & 12 & 0.3911 & 0.0053 \\
{\tt cp}  & planar {\sf S}    &  1  & 8  & 0.4181 & 0.0076 \\
{\tt c8}  & 8 octants         &  8  & 18 & 0.4058 & 0.0174 \\
\hline
$4$-spin  & numerical & $\infty$&        & 0.3509 &
                                          \cite{Stadler:96a} \\
          &TAP   &                  &      & 0.3552 &
                                    \cite{Gross:84}\\
\hline
{\tt LABSP} & open      & $\mathcal{O}(N^2)$ & $\mathcal{O}(N)$
                                           & 0.4634 & 0.0011 \\
{\tt LABSP} & periodic  & $\mathcal{O}(N^2)$ & $\mathcal{O}(N)$
                                           & 0.4713 & 0.0020 \\
\hline
\end{tabular}
\end{center}
\end{table}

The relevant quantity for comparing the simulations with the analytical
theory is the connectivity $z$ of a lattice site $j$, that is, the number
of spins $k\ne j$ such that $J_{jk i_3 i_4}\ne0$ for some $i_3$ and $i_4$.
The value of $z$ depends of course strongly on the interaction pattern, see
Table~\ref{tab:psp}. Note that $z$ is independent of the lattice
dimensionality and of the number $X$ of non-zero interaction strengths per
site.

Numerical simulations were performed by sampling up to $10^8$ spin
configurations from at least $10^5$ different instances (random assignments
of the coupling constants) for each model and values of $N=\prod_{k=1}^d
m_k$ between $8$ and $60$. The standard deviations listed in
Table~\ref{tab:psp} are statistical errors from fitting the curve $\ln
\EE{\mathcal{N}}$ versus $N$ to a straight line. With the exception of the
cubic models, where we have only few data points, and the models {\tt q1q},
{\tt q2q}, which show quite strong finite size effects, the correlation
coefficient of the linear regression is $\varrho>0.9998$.

By comparison with the numerical estimates for the infinite-range 
4-spin glass
we suspect that systematic errors might be slightly larger than the
statistical errors. A conservative estimate appears to be an over all
accuracy of at least $\pm 0.012$. We also note that replacing the Gaussian
distribution of $J_{i_1\ldots i_p}$ by a uniform distribution with mean $0$
apparently does not significantly influence the number of metastable
states.

The results of Table~\ref{tab:psp} show a clear tendency of decrease of
$\alpha$ with increasing connectivities $z$, as expected from the
calculations of section \ref{sect:SR4spin}. Moreover, for fixed $z$ and
within the estimated accuracy, the values of $\alpha$ seem to be
independent of the geometry of the interaction patterns. This interesting
result indicates that the properties of short-range $p$-spin Hamiltonians
may be fully characterized by a few macroscopic parameters, namely, $N$,
$p$ and $z$.  Indeed, the error between our theoretical predictions for
$p=4$ (see Table ~\ref{tab:alpha}) and the numerical data for the
two-dimensional model with the largest connectivity, {\tt qAq}
($z=24$), is only $3.2 \%$. (Comparison of the simulation data with the TAP
solution for $z \rightarrow \infty$ yields an error of $5.4 \%$.)  Since
the estimated systematic error is $1.2 \%$ the agreement between theory and
simulations is quite good and provides additional evidence of the minor
role played by the geometry of the interaction patterns on the landscape
properties of $p$-spin models.
 
\section{The Low Autocorrelated Binary String Problem}
\label{sect:LABSP}

The Low Autocorrelated Binary String Problem ({\tt LABSP})
\cite{Bernasconi:87,Golay:77} consists of finding binary strings $x$
of length $N$ over the alphabet $\{\pm1\}$ with low aperiodic off-peak
autocorrelation $R_k(x) = \sum_{i=1}^{N-k} x_i x_{i+k}$ for all
lags $k$. These strings have technical applications such as the
synchronization in digital communication systems and the modulation of
radar pulses. In the periodic variant of this model, one considers
$R_k(x) = \sum_{i=1}^{N} x_ix_{i+k}$ with indices taken
$\mathrm{mod} N$.
The quality of a string $x$ is measured by the fitness function
\begin{equation}
f(x) = \sum_{k=1}^{N-1} R_k(x)^2\,.
\end{equation}
In most of the literature on the {\tt LABSP} the {\it merit factor}
$F(x)= N^2/(2f(x))$ is used (see \cite{Bernasconi:87}):
using $f$ instead is more convenient for explicit computations.

Recently there has been much interest in frustrated models without explicit
disorder. The {\tt LABSP} and related bit-string problems have served as 
model systems for this avenue of research
\cite{Marinari:94a,Marinari:94b,Migliorini:94}. These investigations have
shown that {\tt LABSP} has a golf-course type landscape structure, which
explains the fact that it has been identified as a particularly hard
optimization problem for heuristic algorithms such as simulated annealing
(see \cite{Militzer:98,Mertens:96} and the references therein). 
In this section we show that {\tt LABSP} has by far more local optima 
than one would expect from its correlation length or interaction order.

We use the fact that every function on the hypercube $\{+1,-1\}^N$ can
be written as a linear combination of the $p$-spin functions
$\varepsilon_{i_1,i_2,\dots,i_p}(x)=x_{i_1}x_{i_2}x_{i_3}\dots x_{i_p}$
to translate {\tt LABSP} into explicit spin glass form.
Explicitly, we have for the aperiodic model \cite{Stadler:96b}
\begin{equation}
f(x) = a_0 + \sum_{k=1}^{\lceil {N\over2}\rceil-1} \sum_{i=1}^{N-2k}
                    2\varepsilon_{i,i+k}(x)
                  + \sum_{k=1}^{N-1}\sum_{i=1}^{N-1}\sum_{j\ne i-k,i,i+k}
                     \varepsilon_{i,i+k,j,j+k}(x) ,
\label{eq:LABSexp}
\end{equation}
where $a_0$ is a constant factor which does not depend on $N$.  A similar
expression can be derived for the periodic model. There are roughly $N^2/4$
non-zero second order contributions and on the order of $N^3$ non-zero
fourth order contributions. Thus the relative weights (amplitudes) of the
$2$-spin and the $4$-spin contributions are $B_2=\mathcal{O}(1/N)$ and
$B_4=1-\mathcal{O}(1/N)$, respectively. For the general definition of $B_p$
see \cite{Stadler:96b}.  The landscape of {\tt LABSP} thus consists of a
(dominant) 4-spin Hamiltonian plus an asymptotically negligible quadratic
component. The correlation length $\ell$ is therefore given by
$\ell=N/8+\mathcal{O}(1)$, which is in excellent agreement with the
estimate $\ell \approx 0.123\cdot N-0.983$ from numerical simulations
\cite{Stadler:92j}.  We note that the generic $4$-spin landscape is
Derrida's $4$-spin Hamiltonian \cite{Derrida:80a} which is a linear
combination of all $\binom{N}{4}$ distinct $4$-spin functions, while
Eq.(\ref{eq:LABSexp}), on the other hand, only contains $\mathcal{O}(N^3)$
non-vanishing $4$-spin contributions. The landscape of the {\tt LABSP} thus
corresponds to a dilute 4-spin glass.

Table~\ref{tab:psp} summarizes a numerical survey of the local minima of
the {\tt LABSP}. For $N\le 70$ we have generated up to $10^8$ spin
configurations at random and checked whether they are local minima.  As
expected, the number of spin configurations that are local minima increases
exponentially as $\exp(\alpha N)$.  The non-exponential pre-factor depends
strongly on $N \mod 4$: separate estimates of $\alpha$ from data for
$N\,\mod\,4=0,1,2,3$ show deviations of up to $\pm0.005$ within each of the
variants of the {\tt LABSP}, while the standard deviations from linear
regression within each data set is almost an order magnitude smaller.  We
conclude that the discrepancy of $0.008$ between the best estimates for
$\alpha_{\mathrm{periodic}}$ and $\alpha_{\mathrm{open}}$, is not
significant, since it is about the same size as the $N\,\mod\,4$
dependence.

\section{Conclusion}
\label{sect:disc}

Metastable states are an important aspect of a landscape's ruggedness.  In
this contribution we have shown that anisotropies, that is, deviations from
a maximum entropy condition, significantly influence the frequency of
metastable states, even when the interaction order $p$ of the spin glass
model, and therefore the correlation length of the resulting landscape, is
kept constant.

The correlation length $\ell$ was introduced as an easy-to-compute measure
of ruggedness. Later on, it turned out that it has desirable algebraic
properties, in particular in the context of Fourier transformation theory
for fitness landscapes \cite{Stadler:96b}. The inability of $\ell$ to
reflect the $z$-dependence of $\langle\mathcal{N}\rangle$ is certainly a
weakness of this measure, in particular when ruggedness measures are used
to predict the performance of optimization heuristics.
 
The anisotropy of short-range spin glass models is the consequence of a
large number of vanishing coupling constants compared with the
corresponding infinite-range $p$-spin model. Following earlier work by
Tanaka and Edwards \cite{Tanaka:80a} we have determined the influence of a
finite connectivity $z$ on the mean number $\EE{\mathcal{N}}$ of metastable
states.  Our result shows that, to first order in $1/z$, the number of
metastable states increases, i.e., short-range spin glasses are in general
more rugged than their infinite-range counterparts. Moreover, the finite
$z$ effect becomes much more pronounced as $p$ increases since the
coefficient of $1/z$ diverges like $p \ln p$.  Of course, large
anisotropies due to very small lattice connectivities cannot be dealt with
by our approach, which is also based on the assumption that all interaction
patterns with fixed connectivity are equivalent.  However, numerical
simulations of several fourth-order spin glasses show that in these cases
$\EE{\mathcal{N}}$ seems to depend only on the lattice connectivity and not
on the detailed geometry of the spin interactions.

The low-autocorrelated binary string problem may be regarded as a strongly
anisotropic 4-spin model, where the interaction strength can assume only
the values $0$ and $1$. This model is a particularly hard optimization
problem \cite{Bernasconi:87,Mertens:96}. In this contribution we have shown
that it exhibits a number of local optima that is by far larger than
expected for a 4-spin model, even taking into account the reduced number of
non-zero coefficients.

Our results indicate that a large class of generic short-range $p$-spin
Hamiltonians may be fully characterized by a few macroscopic parameters,
namely, $N$, $p$ and $z$.  Particular anisotropic (non-generic)
constructions, on the other hand, need not conform to this picture, as the
example of the LABSP shows.

\subsection*{Acknowledgements}

V.M.O.\ thanks Prof.\ P.\ Schuster and Prof.\ P.F.\ Stadler for their kind
hospitality at the Institut f\"ur Theorestische Chemie, where part of her
work was done. The work of J.F.F.\ was supported in part by Conselho
Nacional de Desenvolvimento Cient{\'\i}fico e Tecnol{\'o}gico (CNPq).  The
work of P.F.S.\ was supported in part by the Austrian Fonds zur
F{\"o}rderung der Wissenschaftlichen Forschung, Proj.\ No.\
13093-GEN. V.M.O.\ is supported by FAPESP.

\section*{References}



\end{document}